# Dynamical Mode Recognition of Triple Flickering Buoyant Diffusion Flames: from Physical Space to Phase Space and to Wasserstein Space

Yicheng Chi, Tao Yang, and Peng Zhang[*]

Department of Mechanical Engineering, The Hong Kong Polytechnic University, Hung Hom, Kowloon, Hong Kong

**Abstract**

Triple flickering buoyant diffusion flames in an isosceles triangle arrangement were experimentally studied as a nonlinear dynamical system of coupled oscillators. The objective of the study is two-fold: to establish a well-controlled gas-fuel diffusion flame experiment that can remedy the deficiencies of previous candle-flame experiments, and to develop an objective methodology for dynamical mode recognition that is based on the Wasserstein distance between distribution functions of phase points and can be readily generalized to larger dynamical system consisting of more than three flames. By using the present experiment and methodology, we recognized seven distinct stable dynamical modes, such as in-phase mode, flickering death mode, partial flickering death mode, partial in-phase mode, rotation mode, partial decoupled mode, and decoupled mode. These results generalize the literature results for triple flickering flame system in a straight-line and equal-lateral triangle arrangement. The identification or discrimination of two dynamical modes can be quantified as the small or large Wasserstein distance, respectively.

[*] Corresponding author
E-mail address: pengzhang.zhang@polyu.edu.hk
Fax: (852)23654703, Tel: (852)27666664

**Introduction**

Diffusion flames are ubiquitous in nature (e.g. wildland and urban fires), domestic applications (e.g. fireplaces and furnaces), and industrial applications (e.g. gas-turbine and rocket engines). Vibratory motion of Bunsen-type diffusion flames was observed and referred to as "the flicker of luminous flames" by Chamberlin and Rose[1], who described "the upper portion of the luminous zone rises to a maximum height ten times per second". Subsequently, Barr[2] discovered the similar phenomena in a Burke-Schumann diffusion flame, in which the flame vibrates between the maximum and minimum positions and "the vibration is seen to consists of a progressive necking of the flame which can lead to the formation of a flame bubble which burns itself out separated from the anchored flame". Similar phenomena were also observed in pool fires and often referred to as "puffing flames" [3, 4].

Many studies have been devoted to understanding the physics of flickering diffusion flames. The fact that the flicker of diffusion flames is a self-exciting flow oscillation was substantiated by Chen et al.'s flow visualization of a methane jet diffusion flame [5], in which the small vortices inside of the luminous flame are due to the Kelvin-Helmholtz instability of the fuel jet, and the large toroidal vortices outside the luminous flame are due to the buoyance-induced Kelvin-Helmholtz instability. The frequency of the toroidal vortices was found to well correlate with the flicker frequency [6-13].

In recent years, coupled flickering buoyant diffusion flames as a nonlinear dynamical system have gained increasing attentions. In the pioneer work on dual flickering candle flames by Kitahata et al.[14], two distinct dynamical modes at different flame distances were observed, such as the in-phase mode, in which both flames flicker identically with no phase difference, and the anti-phase mode, in which both flames flicker identically but with a phase-shift of $\pi$. Similar phenomena were also reported by Forrester[15] and Manoj et al.[16] for candle flames. In addition, Manoj et al. [16]observed that an amplitude death mode occurs for small candles and that in-phase and anti-phase modes could coexist and transition to each other. The interaction between buoyance-induced vortices generated by each flame plays a significant role in producing the different dynamical modes at different flame distances. This has been confirmed by the experiments of Dange et al. for candle flames[17], of Fujisawa et al. for pipe-burner diffusion flames[18], by Bunkwang et al. for methane/air jet diffusion flames[19, 20], and by the numerical simulations of Yang et al. for pool flames[21] and of Tokami et al. for buoyance-induced turbulent diffusion flames[22].

Coupled multiple flickering diffusion flames appear more complex dynamical modes. Okamoto et al.[23] investigated three coupled flickering candle flames in an equilateral triangle arrangement. With increasing the size of the triangle, they observed four distinct dynamical modes, such as the in-phase mode, the partial in-phase mode (two of the three flames flicker in phase but the remaining one flickers with a phase-shift of $\pi$), the rotation mode (all three flames flicker with a constant phase-shift of $2\pi/3$), and the death mode (all three flames fall into the stable combustion without any oscillation).

Manoj et al.[24] investigated the coupled four flickering candle flames in a rectangle arrangement and observed the clustering mode, in which the flames separate into two clusters of synchronized flames, the chimera mode, in which the flames separate into synchronized and desynchronized groups of flames, and the weak chimera mode, in which three frequency-synchronized flames coexist with one desynchronized flame. Forrester[15] observed that a ring of flames collectively enhance or suppress the height of a central flame. Manoj et al.[25] experimentally observed very rich dynamical behaviors in a network of flames with various arrangements such as straight line, triangle, square, star, and annular networks. Non-identical asymmetric flames were also studied by Chen et al.[26].

Despite the above noteworthy experimental progress in discovering dynamical behaviors of coupled flickering diffusion flames, there are many interesting problems to be solved and the study is still in its infant stage. Consequently, the present study was motivated by recognizing the following two major deficiencies in the existing experimental studies:

First, most previous experimental studies adopted candle flames, which are experimentally accessible but have a major drawback of the imprecise and inadequate controllability of flame parameters. In these experiments, an observed dynamical mode was often unstable and just sustained for a certain percentage of entire experimental duration. We hypothesized that the lack of adequate flame controllability is responsible for the unstable modes and that a well-controlled gas-fuel diffusion flame experiment could enhance their stability to external disturbances. Very recently, Aravind et al.[27] made an interesting attempt to use an ethanol lamp to enhance the tunability of their flame system.

Second, almost all previous experimental studies reported dynamical modes by presenting their most distinct and representative cases. Based on our experimental results to be expatiated in the paper, many cases significantly deviate from the representative cases and appear as "transition modes" that are very difficult to recognize by observing the time-resolved images and analyzing their flicker frequencies[16, 27]. Dynamical mode recognition in physical space or frequency space becomes increasingly difficult for a system consisting of an increasing number of flames. Inspired by Bifurcation theory, we hypothesized that the different modes are actually caused by the bifurcation of the flame dynamical system due to the change of its parameter values (the bifurcation parameters, e.g. the ratio of flame distance to flame diameter). Consequently, the conceptually correct method for mode recognition would focus on examining the topological structures of phase portraits in phase space.

The present study adopted Bunsen burners to produce flickering buoyant diffusion flames of methane. Owing to the precise controllability of flame parameters and the novel data analysis methodology based on Wasserstein Distance, a concept from optimal transport[28] and deep learning[29] theories, we could recognize more untypical dynamical modes in an objective way. It should be emphasized that, although the present study focused on triple flame systems, the proposed methodologies can be readily extended to larger flame systems.

**Experimental Methodology**

A schematic of the key experimental apparatus established for the present study is shown in Figure 1. Some of the apparatus have been used in our previous experimental works[30-32]. Each Bunsen burner is 1cm in diameter and 12 cm in height. Fuel flow rates are accurately controlled for each individual burner by the MC Series (5SLPM-D/5M) mass flow controller with the range of $m_f = 0 - 5$ slpm (standard liter per minute) produced by Alicat Scientific. Single flickering flame can be produced by a flow rate of 0.45~0.65 slpm. A high-speed camera (Chronos 2.1-HD), which was 2-meter away from the burners, was used to obtain time-resolved (500 fps) shadowgraph images of front views. A fire-proof curtain and mesh screen were used to minimize any external disturbance. As a validation of the present experimental setup, we reproduced the previous experimental results for single and dual flickering flame systems, as shown in the Supplementary Material.

To generalize the previous study of Okamoto et al.[23] on equal-lateral triangle flame system and the study of Manoj et al.[25] on straight-line and equal-lateral triangle flame systems, we arranged the triple flames in an isosceles triangle with variable leg ($L$) and base ($B$). As a limiting case of $B = 2L$, the isosceles triangle is degenerated to a straight line. It will be seen shortly that the D$_2$ symmetry of the isosceles triangle arrangement imposes additional constraints to the flame dynamical system to avoid unnecessary complexity. To further minimize random experimental errors, we ran at least three experiment trials for each flame arrangement, and each experiment trial was continuously recorded for 22 seconds after the flames reached a visually stable combustion state.

**Dynamical Mode Recognition in Physical Space**

To verify our hypothesis that the previously observed unstable modes in candle flame experiments can be stabilized in the present precisely-controlled methane flame system, we examined all the experimental trials and identified seven stable dynamical modes, as shown in Figure 2. These modes can exist within almost the entire time duration of recorded flame videos and were highly repeatable for any longer duration. The emergence of different dynamical modes with varying the flame distances and the fuel flow rate is qualitatively described below by use of the representative cases shown in Figure 2.

If three flame burners are too close to each other, their flames merge into a bigger one with smaller flickering frequency (see Supporting Materials), rendering a case of no interest to the present study. When the flame burners are sufficiently separated (e.g. $B = 4.0$ cm, $L = 2.8$ cm, $m_f = 0.55$ slpm), the in-phase mode (Mode I) appears as the three flames flicker synchronously with negligible phase difference. It is noted that Manoj et al.[25] named the mode as "clustering" where three flame exhibit the same frequency and maintain a constant phase difference. However, they found the case of vanishing phase difference (i.e. in-phase mode) was unstable in their candle experiments.

If flow rate is decreased slightly compared with that in Mode I (e.g. $B = 4.0$ cm, $L = 2.8$ cm, $m_f = 0.45$ slpm), a flickering death mode (Mode II) appears as the three flames oscillate with small amplitude without flicker. The death mode reported by Okamoto et al.[23] is a special case of Model II when the oscillation amplitude is sufficiently suppressed. In addition, furthering the distance of three flame burners, we recognized a new mode (e.g. $B = 5.0$ cm, $L = 3.2$ cm, $m_f = 0.45$ slpm) called partial flickering death mode (Model III), where the two base flames flicker in an anti-phase way while the vertex flame oscillates without flicker.

If the vertex flame is farther compared with that in Mode I (e.g. $B = 4.0$ cm, $L = 4.5$ cm, $m_f = 0.50$ slpm), the partial in-phase model (Mode IV) appears as the two base flames are in-phase to each other but anti-phase to the vertex flame. This mode was reported by Okamoto et al.[23] as an unstable one; Manoj et al.[25] categorized it into a "rotating cluster" mode predominantly observed for the straight-line flame configuration, in which the in-phase flame pairs transition in time.

If the three flames are arranged close to an equal-lateral triangle (e.g. $B = 5.0$ cm, $L = 4.7$ cm, $m_f = 0.50$ slpm), the rotation mode (Mode V) appears as the flames alternatively flicker with a fixed phase difference. The phase difference is $2\pi/3$ for an equal-lateral triangle due to its D$_3$ symmetry, but it can slightly deviate from $2\pi/3$ for the isosceles triangle with D$_2$ symmetry.

If the vertex flame is sufficiently away from the two base flames (e.g. $B = 4.0$ cm, $L = 10.2$ cm, $m_f = 0.50$ slpm), the partial decoupled mode (Mode VI) appears as the two base flames are anti-phase while the vertex flame flickers independently. If the two base flames are also sufficiently away from each other (e.g. $B = 8.0$cm, $L = 10.8$cm, $m_f = 0.50$ slpm), the decoupled mode (Mode VII) appears as the three flames flicker independently. It is noted that Manoj et al.[25] named Mode VI as "weak chimera" and Mode VII as "complete desynchrony".

**Dynamical Mode Recognition in Phase Space**

To verify our hypothesis that different modes of the flame system correspond to the different topological structure of phase portraits in phase space, we proposed to reduce the present infinite-dimensional dynamical system, which is governed by the partial differential equations describing the time-space evolution of the chemically reacting flow, to a finite-dimensional dynamical system, whose temporal evolution can be described by ordinary differential equations. The crucial procedure of the dimension reduction is to choose a certain number of time-dependent variables that can characterize each flickering flames and their interaction. Although there is quite freedom of choice[14-16, 18, 23, 26, 27], we adopted the integral quantity of "flame brightness", which contains certain global information of flame and has been used in some previous studies[14, 23, 26]

On each grayscale high-speed image at a certain time instant, the brightness of each pixel can be represented by an integer from $b = 0$ (pure black) to $b = 255$ (pure white). By using a truncation value of $b = 50$, we obtained a bright contour for each flame. The integration of all brightness values within a flame contour yields a time-dependent scalar quantity $\mathcal{B}_i(t)$, $i = 1 - 3$. Consequently, the dynamical state of the triple flame system at time $t$ is represented by a phase point in a three-dimensional phase space with coordinates $(\mathcal{B}_1(t), \mathcal{B}_2(t), \mathcal{B}_3(t))$. It is noted that the projection effect on the "flame brightness" is negligible because the distance between the camera and the flames is $20 - 100$ times larger than the flame separation distance. The time-evolution of the triple flame system generates a continuous phase trajectory, and the phase trajectory within a sufficiently long-time duration generates an approximately continuous phase portrait, as shown in Figure 3.

The three-dimensional phase portraits and their two-dimensional projections for seven representative stable dynamical modes are shown in Figure 4. It should be noted that all phase portraits have the same time duration (i.e. 22s physical time or about 250 periods) and that all phase spaces have the same ranges of coordinates in all dimensions. Several important observations can be made as follows.

First, the phase volume occupied by a phase portrait can be used as a measure of the amplitude of the change of "flame brightness" due to flicker. As a result, the flickering death mode (Mode II) has the smallest phase volume while the decoupled mode (Mode VII) has the largest phase volume.

Second, the in-phase mode (Mode I) has a 3D phase portrait in the shape of slender ellipsoid, whose major principal axis is along the direction (1,1,1) in the phase space. In the three projection planes, the three 2D phase portraits are all in the shape of slender ellipse, whose major axis is along the direction of (1,1).

Third, the flickering death mode (Mode II) has a 3D phase portrait in the nearly spherical shape and all the three 2D phase portraits in the nearly round disk shape. The partial flickering death mode (Mode III) has a 2D phase portrait in the shape of butterfly, which is a closed loop spreading out along the direction (1, -1) and reflectional symmetric with respect to the axis (1,1). This is a typical phase portrait for two anti-phase flames.

Fourth, the partial in-phase mode (Mode IV) has a 2D phase portrait in the shape of slender ellipse and two 2D phase portraits in the shape of butterfly. This can be understood by that the two vertex-base flame pairs must have the same dynamical behaviors due to the $D_2$ symmetry of isosceles triangle. This symmetry constraint results in the same topological structure of 2D phase portraits on the X-Z and Y-Z planes for all cases, as shown in Figure 4.

Fifth, the rotation mode (Mode V) has three 2D phase portraits in the shape of triangle. It is seen that some phase points deviate from the triangular closed loop due to external disturbance but will be attracted back to the closed loop.

Finally, the decoupled mode (Mode VII) has three 2D phase portraits in the shape of square. All the phase points tend to homogenously spread out in the square probably due to the ergodicity of the decoupled system. The triangular patterns can be barely recognized in the phase portraits as the result of the very weakly flame interaction at large flame separation distance. The partially decoupled mode (Mode VI) has a 2D phase portrait in the shape of butterfly because the two base flames are in anti-phase.

**Dynamical Mode Recognition in Wasserstein Space**

We realized that the above proposed methodology for dynamical mode recognition in phase space lacks sufficient generality and precision. First, many experimental cases do not generate the typical phase portraits as those shown in Figure 5. Furthermore, many cases appear unstable behaviors, in which two or more typical modes transition in time within a certain time duration. Figure 5 shows the phase portraits of some unstable modes U1–U5 as examples. Second, the mode recognition based on the human visual perception of shapes in phase space, is unavoidably subjective and imprecise. In addition, different understanding and cognition usually cause different choices of terminology. Third, the shape of a phase portrait may not be a useful concept in a higher-dimensional phase space, which certainly emerges in a system with more than three flames.

Based on the above considerations and inspired by the Manifold Distribution Hypothesis from Generative Adversarial Networks[29], we proposed to identify different dynamical modes according to the probability distributions of phase points instead of the shapes of phase portraits. Specifically, two dynamical modes are considered the same if the probability distributions of their phase points are close to each other. In mathematics, the Wasserstein distance (or Kantorovish-Rubinstein metric) is a natural way to quantify the closeness of two probability distributions[33]. Being endowed with the Wasserstein distance (metric), the set of probability distributions constitutes a metric space called Wasserstein space.

In the present work, we adopted the simplest 1-Wasserstein distance (a.k.a. the earth mover's distance[34]) defined by

$$W_1(\mu, \nu) = \inf_{\gamma \in \Gamma(\mu,\nu)} \int d(x,y) d\gamma(x,y) \tag{1}$$

where $\mu$ and $\nu$ are two distribution functions on a metric space $(M, d)$, $\Gamma(\mu, \nu)$ is the set of all joint distributions whose marginals are $\mu$ and $\nu$. In the present problem, $M$ is the set of indexed phase points, and $d(x, y)$ is the natural distance between two phase points $x$ and $y$ by counting the difference of their indices. The 1-Wasserstein distance apparently satisfies the three axioms for a metric: 1) $W_1(\mu, \nu) \geq 0$ and the equality holds only for $\mu = \nu$; 2) $W_1(\mu, \nu) = W_1(\nu, \mu)$; and 3) $W_1(\mu, \nu) < W_1(\mu, \gamma) + W_1(\gamma, \nu)$. The details about the theory of Wasserstein distance can be found in [28] and the detailed descriptions about the Matlab code used in the present work for calculating the 1-Wasserstein distance can be found in[35] and will not be repeated here.

To validate the application of 1-Wasserstein distance in the present problem, we selected seven experimental cases, S1–S7, which respectively correspond to Mode I – Mode VII but are different from those cases shown in Figure 4. To facilitate the comparison with the mode recognition based on the shape of phase portraits, we calculated the 1-Wasserstein distance between the corresponding 2D phase portraits. It should be emphasized that Eq. (1) can be readily applied to higher-dimensional phase portraits. As a result, for any two sets of three 2D phase portraits belong to two experimental cases, we obtained a triplet of 1-Wasserstein distances. All the calculation results for the distance between the stable modes S1–S7 (the unstable modes U1–U5) and Mode I – Mode VII are given in Table I. Several rules for mode recognition can be established from the results:

Rule 1: any two cases belonging to the same modes have relatively small distances. It can be seen in Table 1 that the diagonal triplets of $W_1$ are significantly smaller than the off-diagonal triplets. This confirms the recognition of S1–S7 belong to Mode I – Mode VII, respectively. The "smallness" of distance can be determined by a process of "learning" as many as possible cases belonging to the same mode. The Wasserstein distance plays a crucial role in establishing a discriminator in the learning process.

Rule 2: any two cases belonging to different modes have relatively large distances. It can be also seen the relatively large values of $W_1$ for the off-diagonal triplets. Similarly, the "largeness" of distance can be determined by a learning process. Consequently, two cases cannot be treated as the same modes as long as at least one value in the triplet of $W_1$ is relatively large.

Rule 3: any case that has relatively small distances from two or more stable modes should be categorized as unstable mode. For example, the triplet of $W_1$ is (30, 8, 15) between U1 and Mode V, is (64, 8, 13) between U1 and Mode VI, and is (10, 26, 23) between U1 and Mode VII. Physically, this result implies that U1 may appear like any of these stable modes during a certain period. These unstable modes are probably due to the intrinsic dynamical properties of the triplet flame system, which can be described by using bifurcation theory. The unstable modes could also be caused by the structural instability of the system to the external disturbance. The origin and mechanism of the unstable modes merit future experimental and theoretical studies.

**Concluding Remarks**

The present work was motivated by two hypotheses that made by the authors about the existing experimental studies on multiple flickering buoyant diffusion flames as a nonlinear dynamical system of coupled oscillators. The first hypothesis is that the lack of adequate flame controllability is responsible for the unstable modes reported in the previous candle-flame experiments. The second hypothesis is that the different dynamical modes can be discriminated by recognizing their different topological structures of phase portraits in phase space. We successfully verified the two hypotheses in the system of triple flickering buoyant diffusion flames in isosceles triangle arrangement and therefore proposed a new methodological framework for studying dynamical systems of multiple flickering flames.

Bunsen burners were used to produce three identical flickering buoyant diffusion flames of methane, which were precisely controlled by each individual flow rate controller. By minimizing all possible external disturbances, we identified seven distinct stable modes: the in-phase mode, the flickering death mode, the partial flickering death mode, the partial in-phase mode, rotation mode, the partial decoupled mode, and the decoupled mode. These modes can exist within almost the entire time duration (22 seconds) of recorded flame videos and were highly repeatable for any longer duration. These modes include all the previously discovered modes for triple candle flames in a straight-line and equal-lateral triangle arrangement. In addition, we recognized a type of unstable modes, in which two or more stable modes emerge in an experimental trial. In fact, most identified modes in the previous candle-flame experiments are unstable in the present sense. Consequently, the present experiment enhances the structural stability of the dynamical system of coupled flame oscillators.

The proposed new methodology for dynamical mode recognition follows the following procedures. First, the coupled triple flame oscillators constitute an infinite-dimensional dynamical system, which is described by partial differential equations of conservation laws for chemically reacting flows. Second, the infinite-dimensional system can be reduced to a 3D dynamical system by choosing an appropriate characteristic scalar quantity (e.g. the "flame brightness" adopted by the present study) for each flame, Third, the time evolution of the 3D dynamical system generates phase portraits in a 3D phase space, and each phase portrait corresponds to a distribution function. Fourth, the Wasserstein distance quantifies the

"closeness" of two distribution functions so that a small Wasserstein distance indicates the similarity of two phase portraits and the identification of the same dynamical mode. The present calculation results validate the proposed methodology, and three rules for mode recognition were also established and will be further verified in future work.

**Acknowledgement**

This work is financially supported by the National Natural Science Foundation of China (No. 52176134) and by the Hong Kong Polytechnic University (G-UAHP).

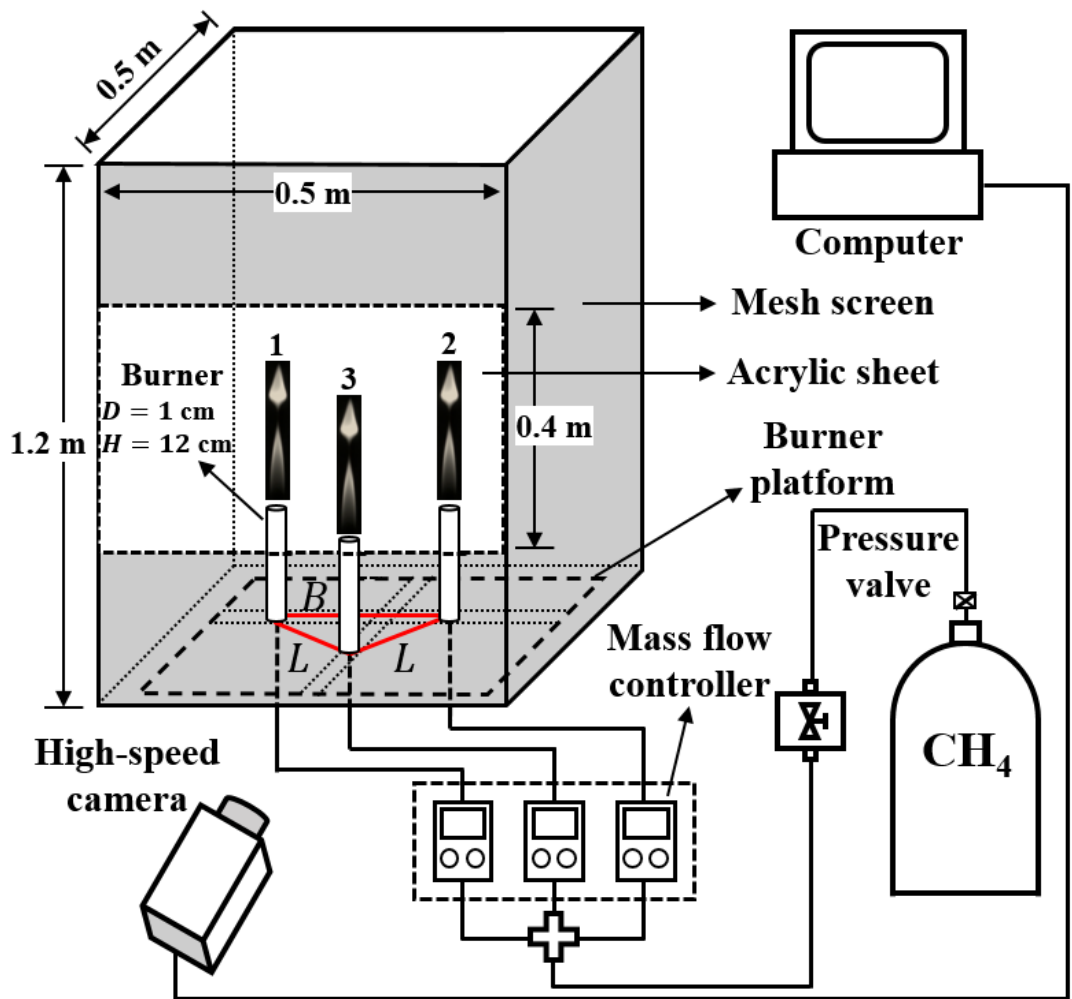

Figure 1 Schematic and photograph of the established experimental apparatus consisting of burners, fuel flow control, and visualization systems.

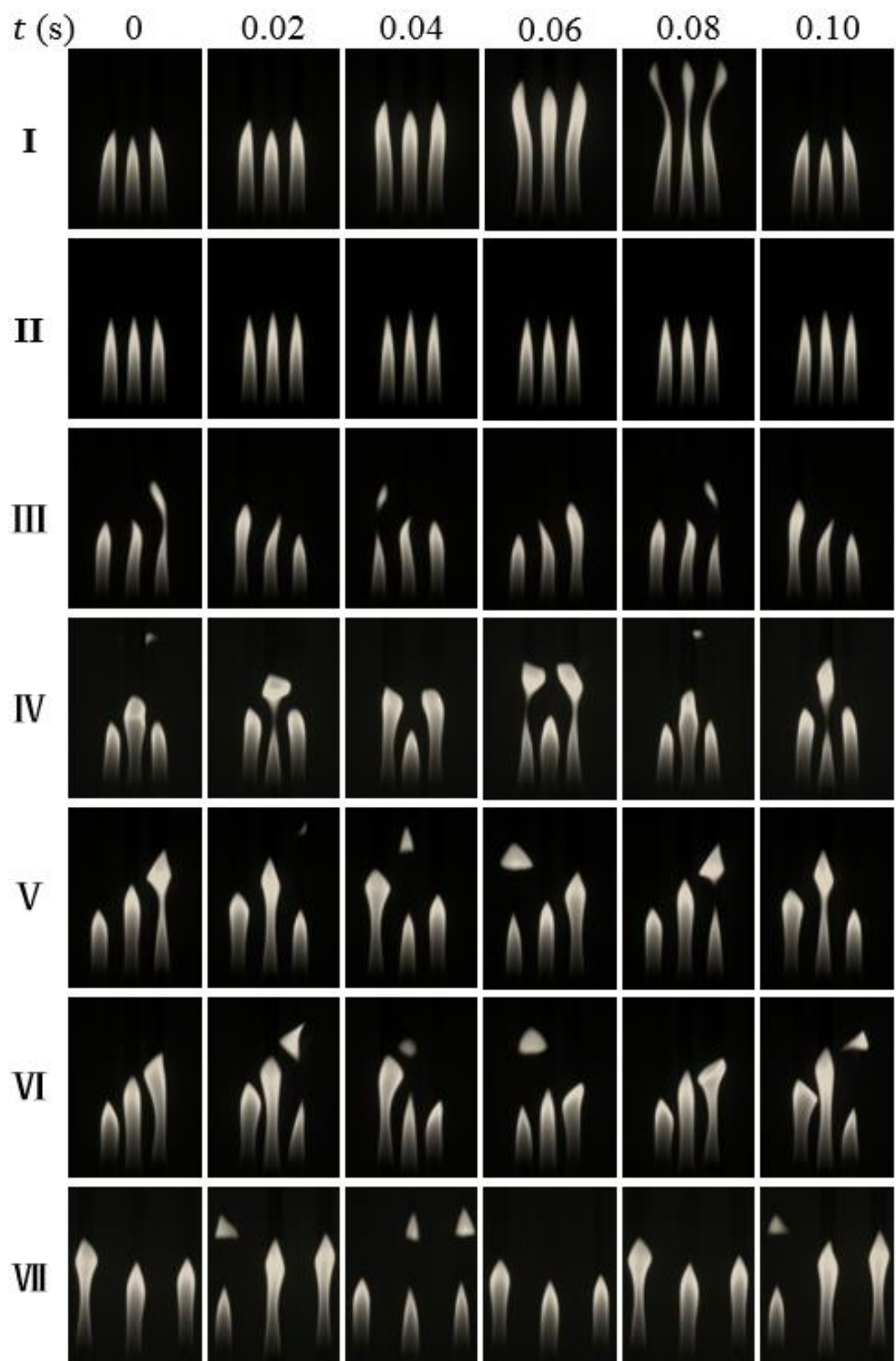

Figure 2. Distinct dynamical modes of triple flickering buoyant diffusion flames of methane. Mode I: In-phase, Mode II: Flickering death, Mode III: Partial flickering death, Mode IV: Partial in-phase, Mode V: Rotation, Mode VI: Partial decoupled, Mode: VII: Decoupled. See Supplementary Materials for the details for the flame parameters.

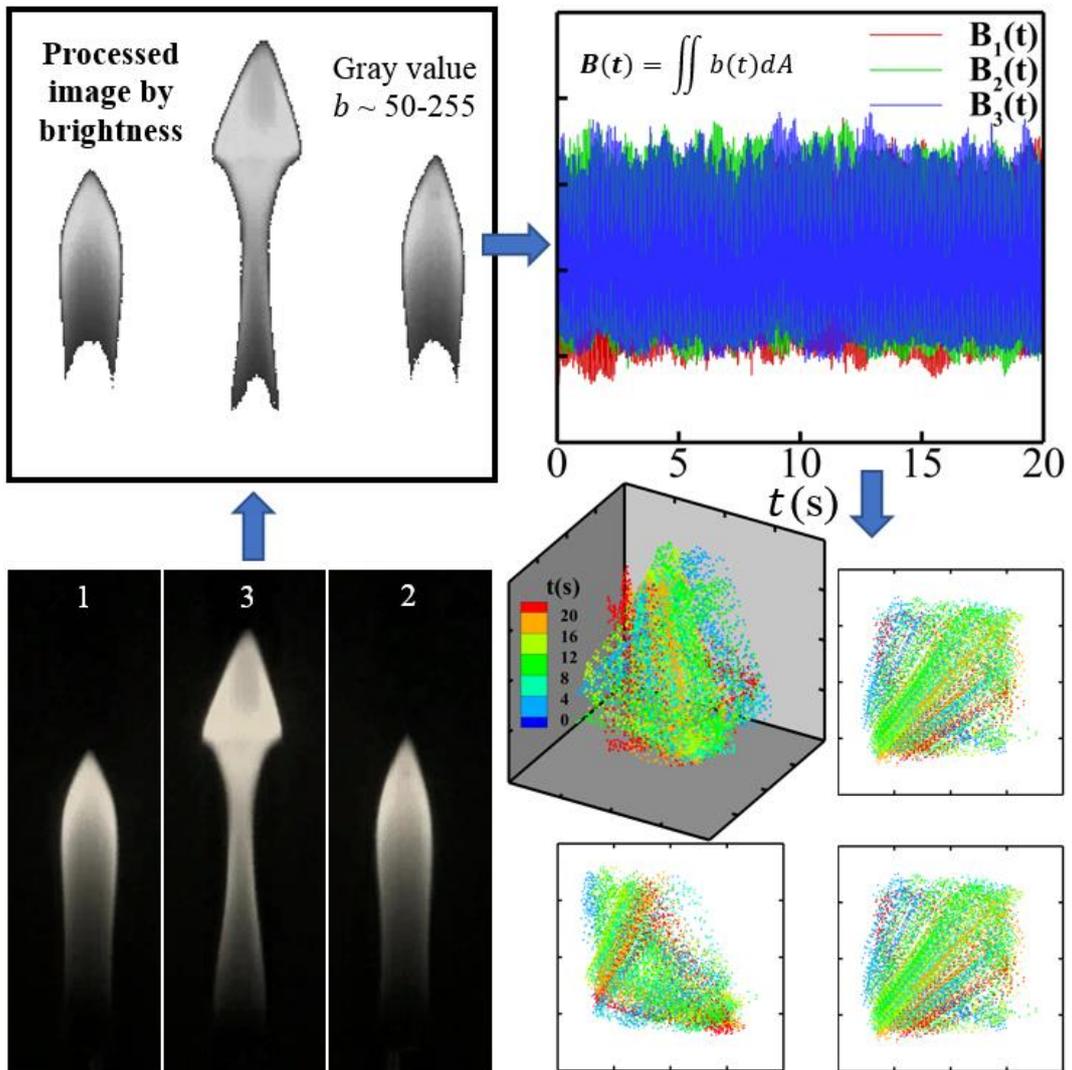

Figure 3. The schematic of establishing phase portraits in a three-dimensional phase space.

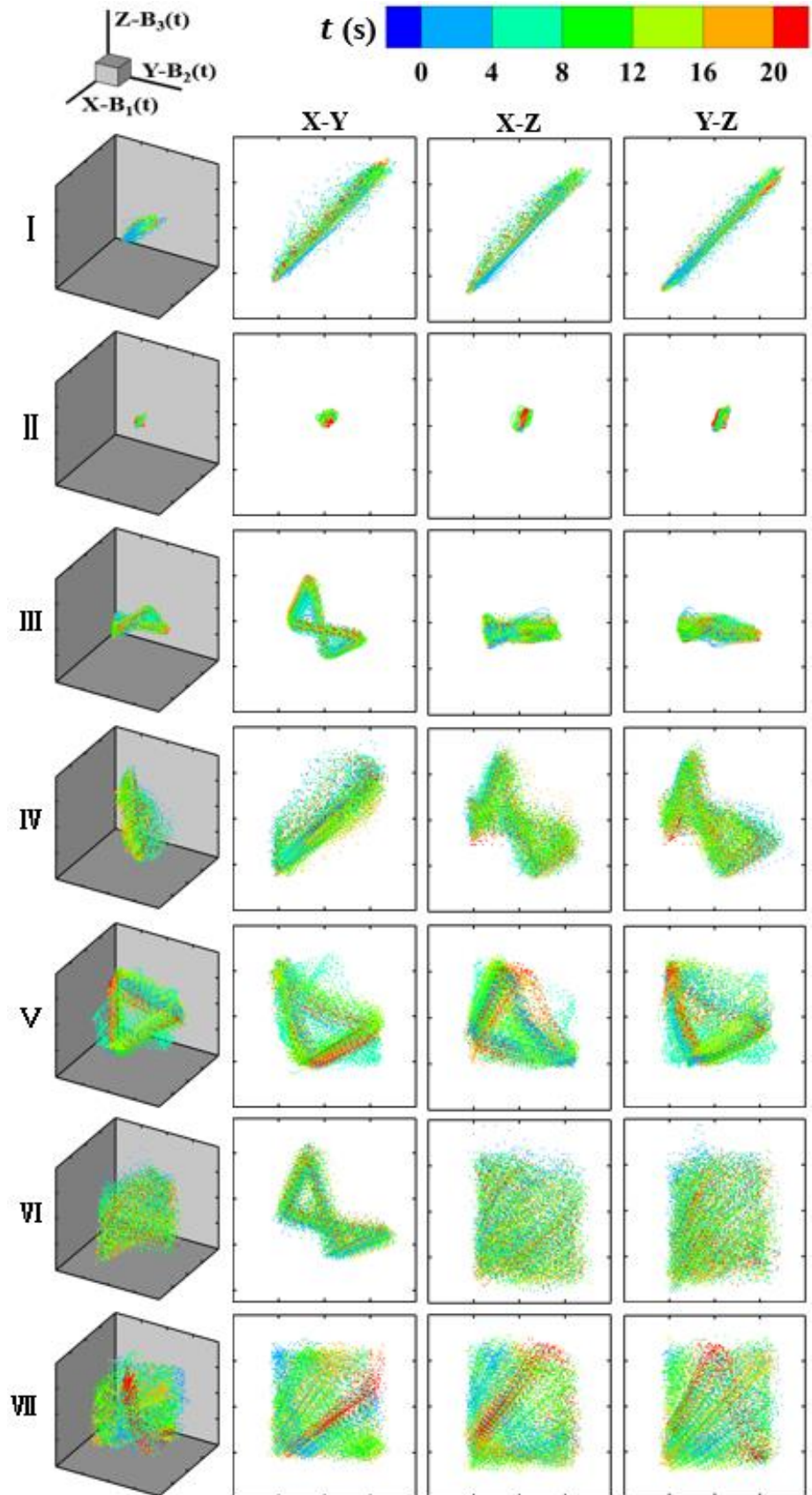

Figure 4. Three-dimensional phase portraits and their two-dimensional projections for seven stable dynamical modes presented in Figure 2. All phase spaces have the same ranges of coordinates in all dimensions.

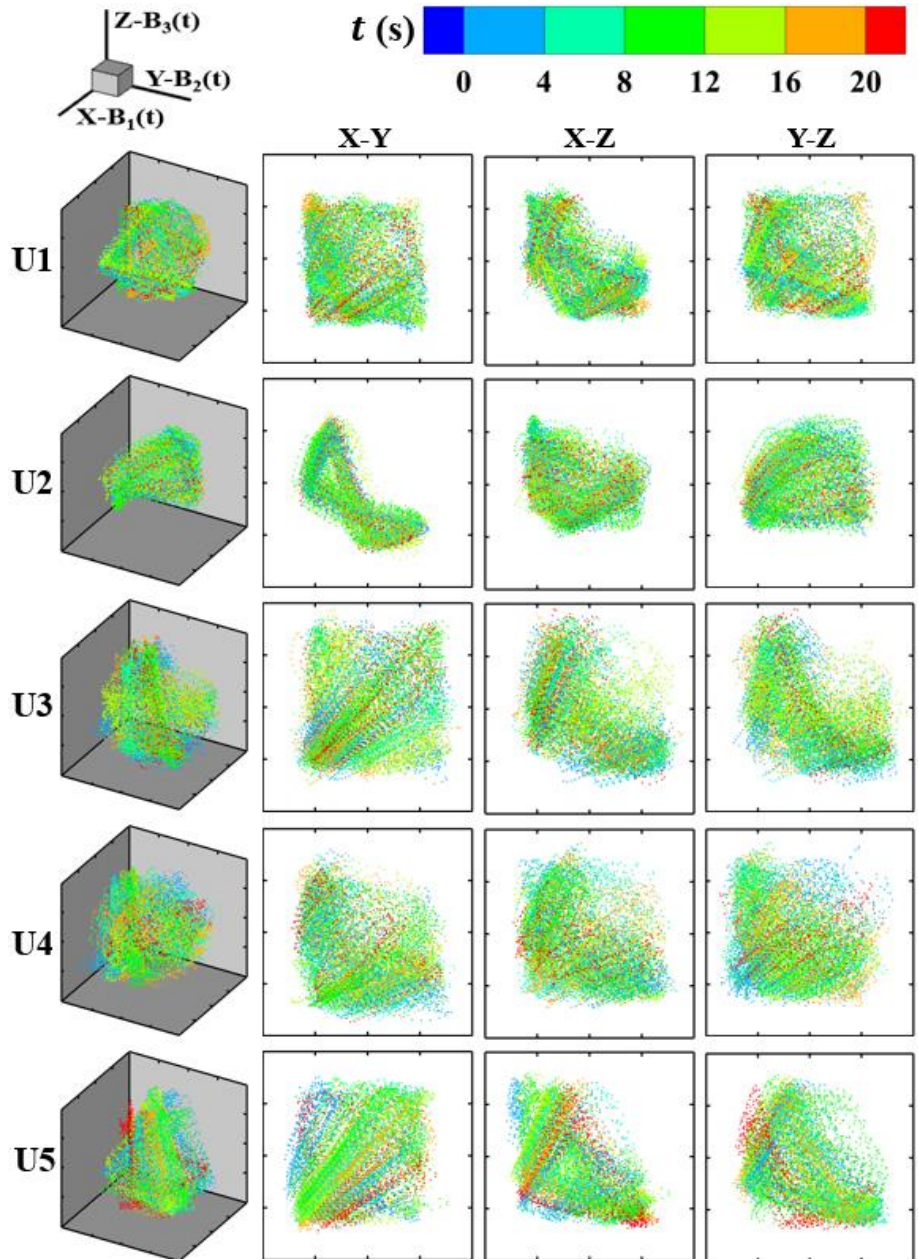

Figure 5 Three-dimensional phase portraits and their two-dimensional projections for unstable dynamical modes. All phase spaces have the same ranges of coordinates in all dimensions.

|   | S1 | S2 | S3 | S4 | S5 | S6 | S7 | U1 | U2 | U3 | U4 | U5 |
|---|---|---|---|---|---|---|---|---|---|---|---|---|
| **I** | 17 | 42 | 84 | 47 | 108 | 62 | 117 | 111 | 61 | 102 | 101 | 99 |
|   | 21 | 53 | 74 | 77 | 100 | 115 | 120 | 108 | 106 | 98 | 109 | 105 |
|   | 27 | 34 | 81 | 77 | 104 | 108 | 120 | 114 | 104 | 103 | 114 | 101 |
| **II** | 26 | 12 | 52 | 23 | 83 | 31 | 92 | 87 | 29 | 78 | 73 | 76 |
|   | 23 | 22 | 50 | 55 | 77 | 95 | 101 | 87 | 84 | 74 | 87 | 83 |
|   | 38 | 25 | 31 | 32 | 62 | 63 | 84 | 73 | 66 | 55 | 70 | 59 |
| **III** | 53 | 31 | 24 | 22 | 62 | 13 | 68 | 63 | 12 | 55 | 45 | 52 |
|   | 69 | 42 | 14 | 14 | 17 | 40 | 48 | 28 | 24 | 13 | 29 | 22 |
|   | 67 | 54 | 9 | 16 | 31 | 33 | 57 | 42 | 37 | 19 | 36 | 28 |
| **IV** | 39 | 21 | 41 | 13 | 68 | 29 | 77 | 71 | 23 | 62 | 56 | 60 |
|   | 66 | 40 | 20 | 9 | 22 | 45 | 53 | 32 | 29 | 21 | 34 | 25 |
|   | 56 | 44 | 16 | 10 | 41 | 42 | 67 | 50 | 47 | 29 | 45 | 37 |
| **V** | 81 | 62 | 29 | 45 | 31 | 47 | 34 | 30 | 43 | 21 | 17 | 21 |
|   | 87 | 66 | 35 | 36 | 15 | 12 | 20 | 8 | 10 | 20 | 12 | 13 |
|   | 86 | 70 | 34 | 40 | 11 | 15 | 26 | 15 | 14 | 27 | 19 | 15 |
| **VI** | 60 | 36 | 18 | 26 | 63 | 11 | 68 | 64 | 10 | 56 | 41 | 54 |
|   | 94 | 73 | 41 | 42 | 19 | 7 | 16 | 8 | 12 | 25 | 8 | 16 |
|   | 90 | 77 | 39 | 42 | 13 | 8 | 22 | 13 | 14 | 22 | 13 | 12 |
| **VII** | 109 | 92 | 60 | 75 | 16 | 78 | 6 | 10 | 77 | 16 | 31 | 18 |
|   | 103 | 86 | 55 | 58 | 35 | 16 | 8 | 26 | 29 | 42 | 23 | 34 |
|   | 103 | 92 | 60 | 66 | 29 | 31 | 8 | 23 | 25 | 46 | 31 | 35 |

Table 1 The calculated triplets of 1-Wassertein distance between the stable modes S1–S7 (unstable modes U1–U5) and Mode I – Mode VII.